\documentclass[doublecol,figures]{epl2}%
\usepackage{amsmath}
\usepackage{amsfonts}
\usepackage{amssymb}
\usepackage{graphicx}%
\shorttitle{Helices at Interfaces}
\shortauthor{Gi-Moon Nam \etal}
\institute{ \inst{1}Institut Charles Sadron, CNRS 23 rue du Loess
67034
Strasbourg cedex 2, France \\
\inst{2} Institute of Fundamental
Physics, Department of Physics, Sejong University, Seoul 143-743,
Korea. \\
\inst{3}Groupe BioPhysStat, Universit\'{e} de Lorraine, 57078
Metz, France }

\abstract{ Helically coiled filaments are a frequent motif in
nature. In situations commonly encountered in experiments coiled
helices are squeezed flat onto two dimensional surfaces. Under
such 2-D confinement helices form "squeelices" - peculiar squeezed
conformations often resembling looped waves, spirals or circles.
Using theory and Monte-Carlo simulations we illuminate here the
mechanics and the unusual statistical mechanics of confined
helices and show that their fluctuations can be understood in
terms of moving and interacting discrete particle-like entities -
the "twist-kinks". We show that confined filaments can thermally
switch between discrete topological twist quantized states, with
some of the states exhibiting dramatically enhanced
circularization probability while others displaying surprising
hyperflexibility. }
\begin{document}

\title{Helices at Interfaces}
\author{Gi-Moon Nam\inst{1,2}, Nam-Kyung Lee\inst{1,2}, Herv\'e Mohrbach\inst{1,3},
Albert Johner\inst{1} and Igor M. Kuli\'{c}\inst{1}}
\maketitle

\section{Introduction}

Helically coiled filaments are found everywhere in living nature. The list of
examples is close to innumerable with the most prominent ones:
FtsZ~\cite{FtsZ}, Mrb~\cite{Mrb}, bacterial flagella~\cite{kamiya,hasegawa},
tropomyosin \cite{Tropomyosin} and intermediate
filaments~\cite{VariousShapesIFs}. More recently microtubules were suggested
to spontaneously form large scale superhelices~\cite{venier, Mohrbach}. Even
whole microorganisms exhibit helicity inherited from their constituent
filaments ~\cite{HelicalBacteria}. The superhelicity of filaments is in some
cases of strong evolutionary benefit as in the example of swimming bacteria
utilizing the rotational motion of their helical flagellar filament for
propulsion~\cite{berg} and tropomyosin's helical "Gestalt-binding" around
actin\cite{Tropomyosin}. In other cases, like for microtubules the purpose of
superhelicity remains so far unknown\cite{Mohrbach}. Paralleling biological
evolution artificial, man made helically coiled structures have been created
including coiled carbon nanotubes~\cite{CoiledHelicalCarbonNanotubes},
DNA\ nanotubes \cite{DNA nanotubes}, coiled helical organic
micelles~\cite{Stupp}.

As happens often in experiments our justified desire to simplify
observation conditions by confining filaments to a \ surface
(coinciding with the focal plane) changes the physical properties
of the underlying objects in initially unanticipated but
physically rather interesting manner. With filament helices being
such a profoundly ubiquitous structure it is the purpose of this
work to investigate the rich physical effects of helical filaments
confinement. As we will show, the confinement changes dramatically
the shape as well as statistical mechanics of the confined helix
generating several notable and surprising effects: a) Enhancement
of cyclisation probability, b) Enhancement of end-to-end
fluctuations and c) Generation of conformational multistability
(despite apparent linearity of constitutive relations). We will
see that the conformational dynamics of confined helices is most
naturally described in terms of discrete particle like entities -
the "twist kinks", cf. Fig. 1. We show that these "twist kinks"
are completely analogous to overdamped Sine-Gordon-kinks from
soliton physics \cite{KinksInSineGordon} as well as loops in
stretched elastic filaments \cite{Loops}. These analogies will
help us to rather intuitively develop a phenomenological
understanding of the underlying physics.

\section{The Phenomenology of Squeezed Helices}

Confined biofilaments throughout literature exhibit often
abnormal, wavy, spiral, and circular shapes that appear not be
rationalized by the conventional Worm Like Chain model. This
riddle of peculiar filament shapes is the starting point of our
investigation. In this letter, we propose a new augmented model of
confined intrinsically curved and twisted chains that leads to a
variety of 2-D shapes matching experimental observations. The
corresponding 2-D geometrical curves we call squeezed helices or
more briefly - squeelices. Filaments are modelled as 2-D confined
Helical Worm Like Chains (cHWLC) with bending modulus $B$, twist
modulus $C$ and a preferred curvature $\omega_{1}$ and twist
$\omega_{3}$.

Phenomenologically (as detailed further below) the main effect of
confinement is to introduce narrow regions where the twist is
highly concentrated and the curvature flips, cf. Fig. 1. These
curvature flip points we will call "twist-kinks" in reference to
the concept of kinks in soliton physics~\cite{KinksInSineGordon}.
Depending on the control parameter
$\gamma\sim\frac{B\omega_{1}^{2}}{C\omega_{3}^{2}}$ two regimes
can be distinguished:

i) $\gamma>1$ : Twist kinks having a positive self-energy are
essentially expelled and can only be thermally activated.

ii) $\gamma<1$ : Twist kinks have a negative self-energy and the ground state
involves a finite density of twist kinks.

In both cases the generic shape motive is a circular arc section
or in special cases closed circles. If present, twist kinks
separate arcs of opposite curvature orientations resulting in a
wavy undulatory 2-D shape. For a squeelix with $\gamma>1$, twist
expulsion results in a circular arc shaped ground state -a feature
that can favor the occurrence of closed filaments, which are very
unlikely in the free helical state (in absence of confinement).

Our study is based on the analytical analysis of the confined \
helical WLC Hamiltonian and numerical Monte Carlo simulations
using the density of states method~\cite{WL_PRL}. The simulations,
primarily designed to illustrate circularization enhancement,
focus on the most curious case of twist expulsion where the ground
state becomes circular up to (weak and localized) edge effects. In
this regime, besides the ground state, the chain will also
comprise excited (wavy looking) states involving a discrete number
of thermally activated twist-kinks. Because an isolated twist-kink
almost freely diffuses along the filament an excited state can be
"hyperflexible" meaning that large shape fluctuations are induced
by simple translational sliding motion of the twist-kink along the
contour. For sets of parameters not resulting in twist expulsion
($\gamma<1$), the ground state itself is periodically wavy. This
case where a squeezed helix forms a wavy periodic 2-D structure
appears less exciting as effects like circularization and
hyper-flexibility will be absent. For this reason we will entirely
focus in this very first study on the more illuminating and
surprising limit $\gamma>1$.

A note for the experimentalist concerning non-ideal filaments (in
real world experiments) seems appropriate here. Bacterial flagella
and microtubules are known to be helical filaments, but their
mechanical characteristics are generally (in experimental
practice) not quite uniform along the filament. There can be
frozen-in lattice defects like a jump in the number of
protofilaments for microtubules~\cite{ChretienDefectsinMTs} or
annealed defects like a boundary between different polymorphic
states in bacterial flagella. In the cases of interest the
mechanical parameters are uniform within some correlation length,
which may even be larger for cleanly prepared filaments than the
contour length under scrutiny. Our model is generic in the sense
that it applies to a uniform sub-filament. A clean observation and
comparison of the shapes described below further supposes that
shape equilibration is possible with uniform mechanical
parameters.
\begin{figure}[h]
\includegraphics*[ width=8.5cm]{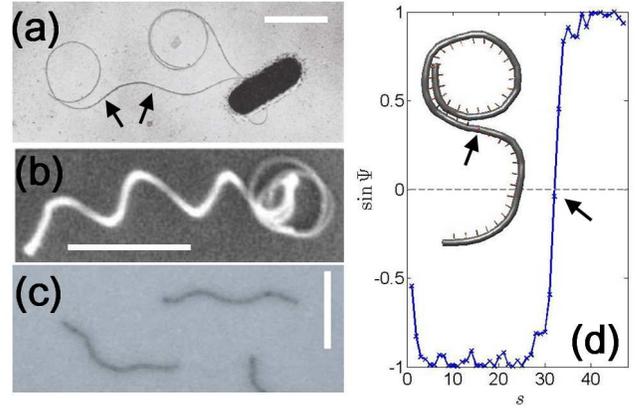} \caption{ Squeezed helical filaments
("squeelices") are frequently observed under experimental conditions. a)
Polymorphic helical bacterial flagella emerging from a bacterium
\cite{asakura} (Scale bar $1.3\mu m$). b) A coiled carbon nanotube when
adsorbed onto the substrate from \cite{CoiledHelicalCarbonNanotubes},(Scale
bar 0.5$\mu m$) c) Helical DNA nanotubes \cite{DNAnanotubes}(Scale bar,
0.5$\mu m$). c) In simulations squeelices appear in coexisting wavy-circular
shapes and display discrete "twist kinks" (curvature inversion points)
indicated by arrows. Twist kinks - discrete non-linear elementary excitations
of confined helices - are characterized by a rapid inversion of the normal
vector along the contour and by a localized twist-angle $\psi$ variation}%
\label{fig:images}%
\end{figure}

\section{The Squeezed Helical Chain Model: Free vs Confined}

The variety of squeelical shapes in Figs. 1 and \ref{fig:FreeEne}
can be understood by simple considerations about their energy. The
elastic energy of a helical wormlike chain can be written as a
function of the local curvatures $\Omega_{1,2}$ and twist
$\Omega_{3}$ as:
\begin{equation}
E=\frac{1}{2}\int\limits_{-L/2}^{L/2}B\left[  (\Omega_{1}-\omega_{1}%
)^{2}+\Omega_{2}^{2}\right]  +C(\Omega_{3}-\omega_{3})^{2}ds, \label{Energy}%
\end{equation}
The 3-D ground state is a helix of radius $R=\frac{\omega_{1}}{\omega_{1}%
^{2}+\omega_{3}^{2}}$ and pitch $H=\frac{2\pi\omega_{3}}{\omega_{1}^{2}%
+\omega_{3}^{2}}$ satisfying the preferred curvature and twist everywhere. To
proceed and access the filament shapes, it is convenient to express the
$\Omega_{i}$ through the Euler angles $\Omega_{1}=\phi^{\prime}\sin\theta
\sin\psi+\theta^{\prime}\cos\psi,$ $\Omega_{2}=\phi^{\prime}\sin\theta\cos
\psi-\theta^{\prime}\sin\psi$ and $\Omega_{3}=\phi^{\prime}\cos\theta
+\psi^{\prime}.$ We constrain the chain to a plane by imposing $\theta=\pi/2$.
The elastic energy of such a confined squeelix can be recast as:
\begin{equation}
E=\frac{B}{2}\int\limits_{-L/2}^{L/2}\left(  (\phi^{\prime})^{2}-2\omega
_{1}\phi^{\prime}\sin\psi+\omega_{1}^{2}+c\left(  \psi^{\prime}-\omega
_{3}\right)  ^{2}\right)  ds \label{Energy2}%
\end{equation}
where $c=C/B$ and $s$ is the arc length parameter along the chain. Now
$\phi^{\prime2}=\Omega_{1}^{2}+\Omega_{2}^{2}\equiv\kappa^{2}$ gives the local
curvature of the chain in the plane and $\psi^{\prime}=\Omega_{3}$ is its
local twist. Interestingly it is seen from Eq.\ref{Energy2} that the curvature
and the twist are now coupled through $\omega_{1}$ - which turns out to lie at
the heart of most phenomena discussed here. The optimal shape satisfies the
Euler-Lagrange equations:
\begin{gather}
\phi^{\prime}=\omega_{1}\sin\psi.\label{Curv}\\
\psi^{\prime\prime}+\frac{\omega_{1}^{2}}{2c}\sin2\psi=0. \label{twist}%
\end{gather}
If no external torque is applied on the chain, the twist must
satisfy the boundary conditions
$\psi^{\prime}(-L/2)=\psi^{\prime}(L/2)=\omega_{3}.$ A quick
glance at Eq. \ref{twist} reveals that $\psi(s)$ is the solution
of a pendulum equation and thus is a Jacobi elliptic function
\cite{elliptic}. Eq. \ref{Curv} shows that under confinement the
curvature becomes "slaved" by the twist and is no more an
independent parameter as for a free chain. This obviously is at
the origin of the localization of twist. For the sake of
simplicity, instead of solving Eqs. \ref{Curv},\ref{twist}
directly, we will gain more physical insight by rewriting Eq.
\ref{Energy2} in terms of a WLC under tension (WLC-T) for which it
is easy to develop intuition. For this we plug in Eq.\ref{Curv}
into the energy Eq.\ref{Energy2} and introduce the new angle
$\vartheta=2\psi-\pi$ to get:
\begin{align}
E\left(  \vartheta\right)   &  =\int\limits_{-L/2}^{L/2}\left(  \frac{1}%
{2}\tilde{A}\vartheta^{\prime2}+\tilde{F}\left(  1-\cos\vartheta\right)
\right)  ds\nonumber\\
&  -\frac{M}{2}\left[  \vartheta\left(  L/2\right)  -\vartheta\left(
-L/2\right)  \right]  , \label{EWLC}%
\end{align}
up to an inessential constant term. Here we introduced $\tilde{A}=C/4$ as the
effective bending modulus of the WLC-T and $\tilde{F}=B\omega_{1}^{2}/4$ as
the effective external force acting on it. The integral term is precisely the
WLC-T energy. The last term with $M=\omega_{3}C$ represents the torque exerted
on both ends of the chain. If large enough, this term enforces $n$ extra
trapped turns/loops $\vartheta\left(  L/2\right)  -\vartheta\left(
-L/2\right)  =2\pi n$ or in the $\psi$ representation, $n$ twist-kinks along
the squeelix.

The phenomenology of the stretched chain is common knowledge
\cite{Loops}, which makes the mapping attractive. The decay length
of any localized distortion (correlation length) $\lambda$ is the
loop size of a loop grown against the tension $F$ (in the loop
picture)\ or kinks-spatial extension (in
the twist-kink picture):%
\begin{equation}
\lambda=\sqrt{\tilde{A}/\tilde{F}}=\frac{1}{\omega_{1}}\sqrt{\frac{C}{B}}%
\end{equation}
The WLC-T\ loop stores a typical energy $\sim\lambda\tilde{F}$. If
the work of the external torques $2\pi M$ per loop reduces
sufficiently the loop energy, loops form spontaneously (otherwise
they can be thermally activated). We hence conclude that the
ground state is wavy for $M\gtrsim\lambda\tilde{F}$, which
translates into $1\gtrsim\frac{B\omega_{1}^{2}}{C\omega_{3}^{2}}$
in the cHWLC language, and circular otherwise.

An overestimate of the energy stored in the loop $2\sqrt{2}\pi\sqrt{\tilde
{A}\tilde{F}}$ is obtained assuming a circular loop. The WLC-T featuring loops
in the long chain limit has been studied in detail in ref. \cite{Loops} from
which we borrow the more precise expression $E_{loop}=8\sqrt{\tilde{A}%
\tilde{F}}$. After subtracting the work of the torques we arrive at the
effective self-energy of a single twist-kink:
\begin{equation}
E_{1kink}=(\sqrt{\gamma}-1)\pi C\omega_{3} \label{twistkinkselfenergy}%
\end{equation}
where
\begin{equation}
\gamma=\frac{4B\omega_{1}^{2}}{\pi^{2}C\omega_{3}^{2}}%
\end{equation}
is the twist-kink expulsion parameter introduced previously. The shape of an
isolated twist-kink is easily obtained upon integrating Eq.\ref{twist}, for a
twist kink localized around $s=0$:
\begin{equation}
\sin\psi\left(  s\right)  =\tanh\frac{s}{\lambda}\quad\text{and}\quad
\phi^{\prime}\left(  s\right)  =\omega_{1}\tanh\frac{s}{\lambda}
\label{twistkinkshape}%
\end{equation}
a result we could also have transposed from \cite{Loops}. From
Eq.\ref{twistkinkshape} it is clear that provided $L\gg\lambda$ the twist kink
is localized and separates two curvature flipped regions of almost constant
curvature $\approx\omega_{1}$ \cite{NOTETwistExpulsion}.

Let us now turn to the main focus of this paper, the case of twist
expulsion ($\gamma>1$), which maps to the ground state to a WLC-T
without loops. In a consistent thermodynamic picture the
twist-kink density should be calculated including shape
fluctuations, in particular the coupling of the twist kink to
small thermal deformations. These linear fluctuations -which in
the soliton picture are often referred to as "phonons"
\cite{KinksInSineGordon} - are expected to only moderately
renormalize the kinks free energy (by $\sim k_{B}T$ per kink). \
Fortunately intuitive reasoning and simple geometry allows us to
calculate filament's shape fluctuations associated with the
\textit{free sliding} motion of the narrow twist-kink along the
filament in the excited state. The associated soft mode (kink's
position) is at the origin of the surprisingly large fluctuations
in the excited state reported in simulations below.

The opposite regime $\gamma<1$ where the squeelix shapes become wavy can be
described along similar lines with the twist kink density being limited by
mutual kink-kink repulsion. Generally speaking for $\gamma<1$ the gas of twist
kinks seizes to be ideal, giving rise to reduced compressibility and in turn
weaker extension fluctuations of the squeelix. It can be shown that the pair
repulsion between twist-kinks decreases with their distance $d$ as
$U_{int}\sim\pi C\omega_{3}\sqrt{\gamma}f(d/\lambda)$ with f$(x)\sim1/x$ for
$x\ll1$ and $f(x)\sim e^{-x/2}$ for $x\gg1$. The dense twist kink regime
deserves special consideration, a more detailed description explicitly
involving Jacobi functions will be given elsewhere.

\section{Simulation}

Equipped with this intuitive picture of a squeezed chain as a
collection of\ weakly interacting 1D "particles" (twist-kinks)
whose positions along the contour relate to chain's deformations
in simple manner (Eq. \ref{twistkinkshape}) we move on to
investigate the formulated hypotheses with a Wang-Landau type
Monte-Carlo simulation \cite{WL_PRL}. We model the system as a
discrete helical WLC consisting of $N$-monomers \cite{langowski}
subjected to a discrete version of the Hamiltonian given in
Eq.\ref{Energy}. The chain is further confined by a harmonic
potential so that each monomer located at distance $z$ away from
the $z=0$ surface experiences potential $E_{conf}=Kz^{2}$ with
$K=25k_{B}T/b^{2}$, with the bond length set as $b=1$. For $K=0$
the confinement vanishes and we recover the free chain statistics
in 3-D.

The main output of the simulation is the density of states (DOS)
\cite{WL_PRL}. The equilibrium conformational statistics can be
obtained from the DOS after Boltzmann weighting according to the
Hamiltonian. In addition to the energy $E$, the joint DOS
$g(E,D,n)$ was sampled with respect to the end-to-end distance $D$
and to the "number of kinks" $n$, where
$n=\frac{1}{\pi}\sum_{i}^{N-2}\Omega_{3,i}$. The latter is indeed
a statistical analogue of the discrete twist kink number in the
theoretical consideration above. Sampling of two-dimensional or
higher histograms is often computationally more demanding than
one-dimensional histograms. For efficient sampling, we used the
global update method introduced by \cite{Zhou_global}. The
simulations were performed for several chain lengths,
$N=16,\,32,\,64$ \ and with fixed mechanical parameters
$B=50,\,C=25\,bk_{B}T$ throughout. The essential information on
the confined cHWLC gained from our simulations are summarized in
free energy maps $F(n,D)$ (see fig.\ref{fig:FreeEne}(b)).

\section{Results}

\begin{figure}[ptb]
\includegraphics*[ width=8.5cm]{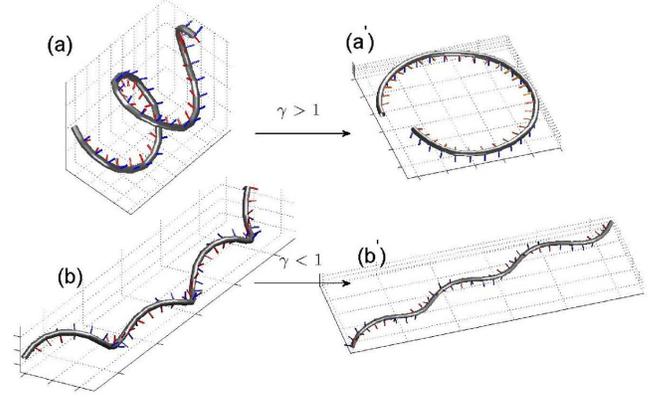}\caption{Formation of two different
classes of conformations of a confined helix depending on the
twist expulsion parameter $\gamma$. The contour is represented by
a tube with a normal (in red) and a binormal vector (in blue). (a)
and (b) typical helical chains in 3-D before confinement. When
confined, (a$^{^{\prime}}$) untwisted circularized shapes
($\gamma>1$) and (b$^{^{\prime}}$) twisted and wavy shapes
($\gamma<1$) emerge.}%
\label{fig:gamma}%
\end{figure}

Figure \ref{fig:gamma} demonstrates the typical changes of chain's
conformation triggered by confinement. For values $\gamma>1$, the confined
helix assumes an approximately circular shape as shown in Fig.\ref{fig:gamma}%
(a). The binormal vectors pointing all in the same direction (in red) indicate
that twist is expelled from the chain. The normal vectors are in-plane and
pointing to the center of circle (in blue). For $\gamma<1$, on the other hand,
the chain assumes a stretched and twisted undulatory wavy shape as seen in
Fig.\ref{fig:gamma}(b). The wavy shape consists of a discrete number of
localized twist kinks as anticipated from theory. At the location of
twist-kinks the in-plane normal vectors and the in plane curvature
$\phi^{\prime}$ flip their signs. The curvature between the twist kinks is
approximately constant $\approx\omega_{1}$ as expected from the theoretical
shape of single twist kinks Eq. \ref{twistkinkshape}.

\bigskip

\subsection{Hypercyclisation}

\begin{figure}[ptb]
\includegraphics*[ width=8.5cm]{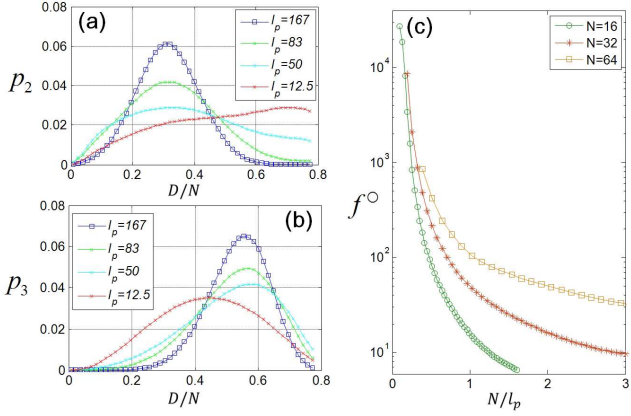} \caption{Confinement enhances
circularization. The end-to-end distance $D$ distribution as a function of
chain's bending persistence length $l_{p}=B/k_{B}T$ (a) in 2D (a) and (b) 3D
($N=32b,$ $\omega_{1}=0.15b^{-1}$, $\omega_{3}=0.1b^{-1}$, and $\gamma
=1.82>1$). The enhancement of circularization probability upon confinement
$f^{\bigcirc}=P_{2d}^{\bigcirc}\left(  D_{cap}\right)  /P_{3d}^{\bigcirc
}\left(  D_{cap}\right)  $ as a function of the normalized chain length
$Nb/l_{p}$ with capture distance $D_{cap}=0.5b$. }%
\label{fig:PD}%
\end{figure}

The phenomenological theoretical arguments above predict that a significant
circularization enhancement should be observable for a twist-expelling chain
($\gamma>1$). We subject this hypothesis to a simulation test. For the
geometric parameters $\omega_{1}N=4.8$ and $\omega_{3}N=3.2$ and elastic
constant ratio $B/C=2$ the twist expulsion parameter is well over unity
$\gamma=1.82$. In absence of confinement, a chain with these parameters
assumes a helical shape with a radius given by $R=0.14Nb$ and a pitch
$H=0.62Nb$. When on the other hand such a chain becomes strongly confined to
2D , it morphs into a slightly open but approximately circular shape with an
end-to-end distance approximately given by $\frac{D_{0}}{Nb}\approx\left\vert
\frac{2}{\omega_{1}N}sin\left(  \frac{\omega_{1}N}{2}\right)  \right\vert
=0.28$. In such a case, the newly formed ground state under confinement seems
to promote circularization. To quantify the circularization enhancement we
define the circularization probability $P_{d}^{\bigcirc}(D_{cap})$ given by
the likelihood to find both chain ends within a fixed (small) capture distance
$D_{cap}$ \newline%
\begin{equation}
P_{d}^{\bigcirc}(D_{cap})=\int_{0}^{D_{cap}}p_{d}(D)dD. \label{eq:Pcap}%
\end{equation}
where $p_{d}\left(  D\right)  =\frac{1}{Z}\int\int g\left(  E,D,n\right)
e^{-E/k_{B}T}dEdn$ ,$\ Z=\int\int\int g\left(  E,D,n\right)  e^{-E/k_{B}%
T}dEdndD$ is the radial probability density for a given
end-distance $D$ and the index $d=2$ or $3$ stands for the
dimensionality of the chains embedding (confined in 2D or free in
3D). Fig.\ref{fig:PD} shows the radial distribution functions
$p_{2}(D)$ (2-D) and $p_{3}(D)$ (3-D) at various temperatures. The
maximum of $p_{3}(D)$ (blue line) is located at about $D=0.55Nb$
close to the expected helical pitch. In contrast the largest
population of the confined
chains is located around the peak of $p_{2}(D)$ (blue line) at $D_{0}%
/Nb\approx0.28$ which corresponds to the expected 2-d circular ground state.
At higher temperature, chain becomes more flexible and the end-to-end
distribution broadens due to the thermal fluctuations.

Integrating the probability density Eq.\ref{eq:Pcap} we can
estimate the circularisation \textit{enhancement factor}
$f^{\bigcirc}=P_{2}^{\bigcirc
}(D_{cap})/P_{3}^{\bigcirc}(D_{cap})$ when the chain is squeezed
to 2D as a function of the chains bending persistence length
$l_{p}=B/k_{B}T$. As suggested on theoretical grounds the closing
probability is greatly enhanced due to the confinement at lower
temperatures (or increased chain stiffness) - with $f^{\bigcirc}$
well in excess of $1000$ (Fig.\ref{fig:PD}c).

\subsection{Multistability and Hyperflexibilty}

The theoretical considerations of multi-kink solutions presented above
indicate that confining a helical chain should generate a complex energy
landscape with many coexisting discrete states. Some of these states, in
particular those comprising only a few kinks, should exhibit anomalous
hyperflexible behavior due to the energetically cheap displacement of kinks.
For instance the motion of a single kink on a chain of length $L$ and
curvature $\omega_{1}$ satisfying a close to "resonance" condition $\omega
_{1}\approx k\frac{2\pi}{L}$ ($k=1,2...$ )\ should give rise to most dramatic
end fluctuation effects.

\begin{figure}[tb]
\includegraphics*[ width=8.5cm]{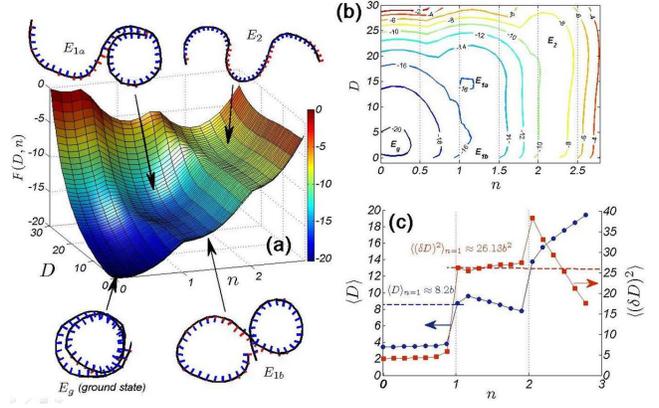} {}\caption{ (a)The free energy
landscape of a fluctuating squeelix as a function of the end-to-end distance
$D$ and the number of twist kinks $n$ ($N=48b$ , $\omega_{1}=0.26b^{-1}$,
$\omega_{3}=0.1b^{-1}$, and $\gamma=5.47>1$). The ground state $E_{g}$ ($n=0$)
is a two-turn circular shape without kinks. The excited states $n\geq1$
comprise one or more very mobile kinks giving rose to strong shape
fluctuations with states $E_{1a}$ and $E_{1b}$ having identical energies. (d)
Contour plot of free energy for 2-d chain with indicated values free energy
values in units of $k_{B}T$. (c) Mean of end-to-end distance $\langle
D\rangle$ and its standard deviation $\langle\left(  \delta D\right)
^{2}\rangle$ as a function of $n$ (dual axis graph).}%
\label{fig:FreeEne}%
\end{figure}

To test this we calculate the joint DOS of three variables
$g(E,D,n)$ and the free energy landscape given by $\beta
F(D,n)=-ln\left[ \sum_{E}g(E,D,n)e^{-\beta E}\right] $ (Fig.
4a,b). For parameters $\omega_{1}=0.26b^{-1}$ and
$\omega_{3}=0.1b^{-1}$, in 3D\ (confinement free case) we would
obtain a helix with helical radius $R=8.09b$ and pitch $H=3.35b$.
In contrast to that as expected from the large twist expulsion
parameter $\gamma=5.47$ the chain's 2D ground state shape under
confinement becomes a coiled double circle with roughly two turns.
Besides the ground state, the free-energy as a function of $n$
indicates the existence of further ("excited") low energy states
with $n=1,2,...$ . (Fig.~\ref{fig:FreeEne}). These states with
distinct $n$ are separated by small free energy barriers. The
first excited state $n=1$ for instance has a free energy
difference of only $\Delta G_{1kink}\approx4k_{B}T$ a figure that
is not far from the theoretical estimate of the kink free energy
$\Delta G_{1kink}=E_{1kink}-k_{B}T\ln N\approx6.6k_{B}T$ \ where
the $\ln N$ (N=48) term accounts for kinks positional entropy gain
along the discrete positions of the chain \cite{NOTEHelix}.

The $n=1$ state (cf. Fig.~\ref{fig:FreeEne}(c)) exhibits an
enhanced end-end distance fluctuation $\langle\left(  \delta
D\right)  ^{2}\rangle$ in phenomenological agreement with the
mobile kink interpretation. For the given length ($L=48b$) and
curvature we expect an almost flat free energy landscape as a
function function the end-end distance $D$ in the range
$D\in\left[ 0,D_{\max}\right]  .$ Here$\
D_{\max}\approx4\omega_{1}^{-1}=15.\,\allowbreak 4b$ is the
maximal extension for the $n=1$ state which is reached when the
twist-kink is located at a position $\approx L/4$ from any of the
borders- as seen from straight forward geometric reasoning. The
standard deviation expected in this case $\langle\left(  \delta
D\right)  ^{2}\rangle_{n=1}$ $\approx$ $28b^{2}$ compares well
with the simulation result $\langle\left( \delta D\right)
^{2}\rangle_{n=1}$ $\approx26b^{2}$.

The higher states $n\geq2$ display again lower fluctuations. This is in
agreement with the interpretation that with growing kink density their
repulsion and eventually mutual confinement become important.

\section{Conclusion}

We have shown that the conceptually simple procedure of planar
confinement transforms a simple mundane object - a helically
coiled filament- into a complex multistable, anomalously
fluctuating filament. The statistical mechanics of this exotic
object -the "squeelix"- can be qualitatively understood in terms
of the motion of discrete particle-like entities corresponding to
sharp curvature inversion points - called "twist kinks". At low
twist kink concentrations they move almost freely along the chain
and induce anomalously strong conformational fluctuations- notably
deviating from wormlike chain behavior. The "squeelical" shapes
formed under confinement range from almost ideally circular to
wavy depending on the value of a single dimensionless "twist
expulsion" parameter $\gamma=\frac
{4B\omega_{1}^{2}}{\pi^{2}C\omega_{3}^{2}}.$ The latter parameter
depends both on the chain's elastic moduli (flexural modulus $B$,
twist modulus $C$) and the geometric properties (intrinsic
curvature $\omega_{1}$ and twist $\omega_{3}$). For $\gamma>1$,
the twist is curiously expelled from the chain. Under these
conditions the squeelix becomes almost circular (up to minor
end-effects) and the circularization probability can be
dramatically enhanced. In the other limit $\gamma<1$ the squeelix
comprises densely packed twist-kinks in its ground state which is
now wavy.

We may speculate whether abnormal wavy shapes or enhanced closure
of confined filaments found in literature are fingerprints of an
otherwise hidden helical superstructures whose microscopic origin
should be elucidated for each type of filament, case by case. Such
peculiar behavior under confinement is observed for essential
filaments like microtubules \cite{venier}, F-actin \cite{Sanchez}
and possibly for intermediate filaments \cite{KoesterPRLonIF}. It
has not escaped our attention that actin filaments circularize on
stunningly small scales ($\sim $5$\mu m$ length rings)
\cite{Sanchez},\cite{Janmey} and exhibit wavy periodic
tangent-correlation functions in narrow, flat -channels (cf. Fig
.6 in \cite{KoesterActinWiggles}) - both phenomena that neither
can be understood within the naive WLC model. Similarly
intermediate filaments under confinement show wiggly periodic
shapes \cite{KoesterPRLonIF} suspiciously reminiscent of squeezed
helices.

It is our feeling that the here illustrated phenomena of multistability,
hyperflexibility and enhanced chain circularization are just the tip of an
ice-berg and that floppy "squeelix" concept will help us to unravel a number
of mysteries from biofilament realm\ hidden out in literature. We have
evidence that additional lateral confinement of filaments in narrow channels,
as encountered in microfluidic devices \cite{KoesterPRLonIF}, further enhances
the visibly wavy shapes for helical filaments and gives easier access to the
underlying mechanical parameters. The potential of the 2-D and especially
double confinement experiments (in microchannels) has been vastly
underestimated so far. The "squeelix" phenomenology laid out here will serve
us as a "dictionary" to decode these peculiar observations in forthcoming works.

\textbf{Acknowledgment} G.-M. Nam and N.-K. Lee acknowledge
financial support of Korean Research Foundation via Grants NRF
2009-0084933 and NRF 2008-314-C00155. I.M.K. and A.J. acknowledge
support from STAR exchange program and insightful discussions with
Sarah K\"{o}ster, Daniel Riveline and Albrecht Ott.


\begin{thebibliography}{99}
\bibitem {elliptic}The solution is $\psi\left(  s\right)  =am\left(  \sqrt
{a}\left(  s-s_{0}\right)  |\frac{\omega_{1}^{2}}{ca}\right)  $ (elliptic
amplitude function) where the constants of integration $a$ and $s_{0}$ are
determined by the boundary conditions.
\bibitem {NOTEHelix}In contrast the corresponding 3-D (free chain's) free
energy appears rather tame and featureless, displaying no multiple
local minima or barriers and lacking the complexity of its 2-D
confined chain counterpart (data not shown).
\end{thebibliography}
\end{document}